\documentclass[%
reprint,
 amsmath,amssymb,
 aps,
 prl,
 rmp,
 prstab,
 prstper,
 floatfix,
]{revtex4-1}
\usepackage{siunitx}
\usepackage{graphicx}
\usepackage{dcolumn}
\usepackage{bm}
\usepackage{tabu}
\usepackage[flushleft]{threeparttable}

\usepackage{xcolor}
\usepackage{tikz}
\usetikzlibrary{shapes}
\definecolor{blue}{RGB}{0,112,192}
\definecolor{lightblue}{RGB}{0,176,240}
\definecolor{green}{RGB}{0,176,80}
\definecolor{yellow}{RGB}{255,255,0}
\definecolor{orange}{RGB}{255,192,0}
\definecolor{red}{RGB}{255,0,0}
\definecolor{darkred}{RGB}{118,0,0}
\definecolor{purple}{RGB}{208,0,154}

\newcommand*{\tikzcirc}[2]{%
   \setbox0=\hbox{\strut}%
   \begin{tikzpicture}
     \useasboundingbox (-.25em,0) rectangle (.25em,\ht0);
     \filldraw[draw=#1,fill=#2] (0.0,0.35\ht0) circle[radius=.25em];
   \end{tikzpicture}%
}
\newcommand*{\tikzrect}[2]{%
   \setbox0=\hbox{\strut}%
   \begin{tikzpicture}
     \useasboundingbox (-.25em,0) rectangle (.25em,\ht0);
     \filldraw[draw=#1,fill=#2] (-0.25em,0.05em) rectangle (.25em,0.55em);
   \end{tikzpicture}%
}
\newcommand*{\tikzdiam}[2]{%
   \setbox0=\hbox{\strut}%
   \begin{tikzpicture}[rotate = 45]
     \useasboundingbox (-.25em,0) rectangle (.25em,\ht0);
     \filldraw[draw=#1,fill=#2] (-0.25em,0.0em) rectangle (.25em,0.5em);  
   \end{tikzpicture}%
}
\begin{document}

\preprint{APS/123-QED}

\title{Rheology of Thickly-Coated Granular-Fluid Systems}

\author{Teng Man$^1$}
\author{Qingfeng Feng$^2$}
\author{K.\ M.\ Hill$^1$}
\email{kmhill@umn.edu}
\affiliation{$^1$Department of Civil, Environmental, and Geo- Engineering, University of Minnesota, Minneapolis, Minnesota, USA.}
\affiliation{$^2$State Key Laboratory of Hydroscience and Engineering, Tsinghua University, Beijing, China}

\date{\today}

\begin{abstract} 
We investigate the link between particle-scale dynamics and bulk behaviors of thickly-coated particle-fluid flows using computational simulations. We find that, similar to dense fully-saturated slurries, the form the rheology takes in these systems can carry signatures of interparticle collisions and/or interparticle viscous dynamics that vary solid fraction. However, we find significant qualitative and quantitative differences in the transitions of the system between what might be called viscous, collisional, and visco-collisional behaviors.  We show how these transitions arise from changes in the ``fabric,'' e.g. strong force network, and the ``granular temperature,'' or fluctuation energy, and suggest extension of these frameworks to better elucidate other particle-fluid behaviors.
\begin{description}
\item[Usage]
Secondary publications and information retrieval purposes.
\item[PACS numbers]
47.57.Gc, 81.05.Rm, 83.10.Pp, 83.80.Hj
\item[Structure]
You may use the \texttt{description} environment to structure your abstract;
\end{description}
\end{abstract}

\maketitle


Thickly-coated particle-fluid flows are ubiquitous in natural and man-made structures, from muddy geophysical flows to concrete and other construction materials. Yet our physics-based understanding of these flows is lacking. The complexity of modeling these flows lies in part in mesoscale ephemeral structures that influence internal resistance to flow. In the last decade, significant progress has been made in modeling closely related systems -- dry particle flows  \cite{pouliquen2002,midi2004,pouliquen2006,jop2006}, suspensions \cite{boyer2011}, and slurries  \cite{cassar2005,trulsson2012} -- by considering the rheology in the context of a non-dimensionalizing shear rate $\dot \gamma$ using a local, collisional or viscous, timescale: 

\begin{subequations}\label{eq:1}
\begin{minipage}{0.225\textwidth}\begin{align}
\label{eq:1a}
I_c=\frac{\dot\gamma}{\sqrt{\sigma/(\rho_{p} d^2)}}
\end{align}
\end{minipage}
\begin{minipage}{0.225\textwidth}\begin{align}
\label{eq:1b}
I_v=\frac{\dot\gamma}{ \sigma/\eta}
\end{align}
\end{minipage}
\end{subequations}
Here, $\sigma$ is the normal stress; $d$ and $\rho_p$ are the particle size and material density; $\eta$ is the dynamic fluid viscosity. What we label here as $I_c$ and $I_v$, are typically referred to as \textit{inertial numbers} and are also often expressed as stress ratios. $I_c$ takes a form proportional to the square root of Bagnold's classic dispersive stress\cite{bagnold1954}, essentially an interparticle collisional stress ($\tau_c \sim \rho_p (d\dot\gamma)^2$) divided by $\sigma$; $I_v$ is a viscous stress ($\tau_v \sim \eta \dot\gamma$) divided by $\sigma$.  Simulations and experiments \cite{pouliquen2002,midi2004,pouliquen2006,jop2006,boyer2011,cassar2005} have demonstrated that behaviors of a wide range of wet and dry particle flows can be efficiently expressed in terms of the dependence of $\mu_{\textrm{eff}} \equiv \tau / \sigma$, and solid fraction $\phi$ on the relevant inertial number $I_c$ and/or $I_v$ (e.g., Table \ref{table:1}).

Trulsson et al.\ \cite{trulsson2012} built on this previous work by hypothesizing that the shear stress ($\tau$) in dense slurries may be expressed in terms of a linear superposition of contact and viscous stresses scaled with a single function of solid fraction $f(\phi)$: 
\begin{equation} \label{eq:2}
\tau=f(\phi)\times(\alpha \rho_p (d \dot\gamma)^2 + \eta\dot\gamma)
\end{equation}
$\alpha$ (a constant) and $f(\phi)$ are empirically determined.  We note that the significance of this suggestion lies in part on the implication that the relative contribution to internal resistance in particle-fluid systems by interparticle contacts (e.g., via $\tau_c$) and fluid-mediated interactions (e.g., via $\tau_v$) is independent of $\phi$.   
Trulsson et al.\ \cite{trulsson2012} showed that using Eqn.\ (2) they could express $\mu_{\rm{eff}}$ in terms of a superposition of inertial numbers:
\begin{subequations} \label{eq:3}
    \begin{align}
        \mu_{\textrm{eff}}= \tau_s / \sigma =  f(\phi)\times I_s; \label{eq:3a} \\
        I_s \equiv \alpha I^2_c + I_v \label{eq:3b}
    \end{align}
\end{subequations}
They found the dependence of $\phi$ and $\mu_{\textrm{eff}}$ on $I_s$ in slurries to be analogous to those previously obtained for particle-fluid flows dominated by $I_c$ or $I_v$ alone (Table \ref{table:1}).

In this Letter, we computationally model behaviors of thickly-coated particle-fluid systems. We find that their rheology can be expressed using similar constructs as for fully saturated flows. Specifically, we find somewhat analogous contributions to the rheology by contact and viscous particle-scale interactions previously seen in slurries \cite{trulsson2012}. However, in our coated systems, we find that there are multiple transitions, or dynamic pathways, between collisional, viscous, and visco-collisional behaviors that are mediated by particle concentration and variations in strong force networks. 

To study thickly-coated particle-fluid systems, we use 3-d discrete element method (DEM) simulations \cite{cundall1979}. Our particles interact via model \textit{contact} forces whenever particle surfaces touch and via model viscous \textit{liquid} forces whenever particle surfaces are sufficiently close. The interparticle contact forces depend on particle properties according to Hertz-Mindlin contact theories, Coulomb friction laws, and a damping component as described in Refs. \cite{tsuji1992,yohannes2010,hill2011}. To model the effect of the viscosity of the coating, we treat the fluid as one that moves with the particles, rather than in a separate phase, i.e., in the form of a lubrication model representing the viscous drag force $\vec{F}_{v}^{\textrm{ij}}$ on interparticle movement \cite{pitois2000}: 
\begin{subequations} \label{eq:4}
\begin{align}
F_{v}^{\textrm{ij,n}} = 6\pi\eta R_{\textrm{eff}}^{2}G_{f}^{2}\frac{v_{n}^{\textrm{rel}}}{\delta_g} \label{eq:4a} \\
F_{v}^{\textrm{ij,t}} = 6\pi\eta R_{\textrm{eff}}v_{t}^{\textrm{rel}}\left[\frac{8}{15}\textrm{ln}(R_{\textrm{eff}}/\delta_g)+0.9588\right] \label{eq:4b}
\end{align}
\end{subequations}
$F_{v}^{\textrm{ij,n}}$ and $F_{v}^{\textrm{ij,t}}$ are normal and tangential components, respectively, to the plane of contact between particles $i$ and $j$; $R_{\textrm{eff}}=[2(d_i^{-1}+d_j^{-1})]^{-1}$; $d_i, d_j$ are contacting particle diameters; $\delta_g$ is the smallest distance between neighboring particle surfaces; $v_{n}^{\textrm{rel}}$ and $v_{t}^{\textrm{rel}}$ are the normal and tangential velocities of these neighboring surfaces. $G_{f}$ and the bracketed term represent non-infinite fluid volume effects. We use regularizing length scales of $\delta_{\textrm{g,min}}\sim \langle d \rangle /10$ and $\delta_{\textrm{g,max}} \sim \langle d \rangle$ (see Supplemental Material I).  Table \ref{table:2} provides the range of particle and fluid properties we used. In each simulation we sheared 6400 particles.

\begin{table}[ht]
\begin{threeparttable}
\caption{Fits and fit parameters for data in Fig. 1}
\centering 
\begin{tabular}{|l|c|c|} 
\hline\hline        
 & $\mu_{\rm{eff}}$ & $\phi$ \\ [1ex] 
\hline                  
 dry & $ \mu_{\rm{1c}}+\frac{\mu_{\rm{2c}}-\mu_{\rm{1c}}}{1 + I_{co}/I_c}$ & $\phi_m\times\frac{1}{1+\beta_c I_c}$ \\ [1ex] 
    beads \cite{pouliquen2002} & $\mu_{1c}, \mu_{2c}, I_{co}=0.27, 0.63, 0.35$ & $\phi_m, \beta_c=0.61,0.25$\\ [1ex]
\hline                  
suspen- & $\mu_{\rm{1v}}+\frac{\mu_{\rm{2v}}-\mu_{\rm{1v}}}{1 + I_{vo}/I_v}+I_v+\frac{5}{2}\phi \sqrt{I_v}$ & $\phi_m\times\frac{1}{1+\beta_v \sqrt{I_v}}$\\                        
sions \cite{boyer2011} & $\mu_{1v}, \mu_{2v}, I_{vo}=0.27, 0.63, 0.05$ & $\phi_m, \beta_v=0.61,0.83$\\ [1ex]
\hline                  
slurries\cite{trulsson2012} & $ \mu_{\rm{1s}}+\frac{\mu_{\rm{2s}}-\mu_{\rm{1s}}}{1 +\sqrt{I_{so}/I_s}}$ & $\phi_m - \beta_{s1} \sqrt{I_s}$ \\ [1ex]  
& $\mu_{1s}, \mu_{2s}, I_{so}=0.27, 2.2, 0.25$ & $\phi_m, \beta_{s1}=0.61,0.40$\\ [1ex]
\hline\hline
\end{tabular}
\label{table:1}
\end{threeparttable}
\end{table}

\begin{table}[ht]
\begin{threeparttable}
\caption{Input Parameters}
\centering 
\begin{tabular}{c c c c c c}
\hline\hline        
$\langle d\rangle \pm \sigma_d $ & $\rho_p $ & $\eta$ & $\sigma$ & $ u_h$ & $\dot{\gamma}=u_h / h$  \\ [0.5ex]
(mm) & ($kg/m^3$) & $(cP)$ & $(Pa)$ & (m/s) & $(1/s)$  \\ [0.5ex]
\hline                  
$1\frac{1}{4} \pm \frac{1}{4}$ & 2650 & $ 10^{- 4}-10^{4}$ & $100 - 500$ & $10 - 10^{3}$ & $0.16 - 46 $ \\ [1ex] 
\hline\hline
\end{tabular}
\label{table:2}
\end{threeparttable}
\end{table}

\begin{figure}[h!]
  \includegraphics[scale = 0.35]{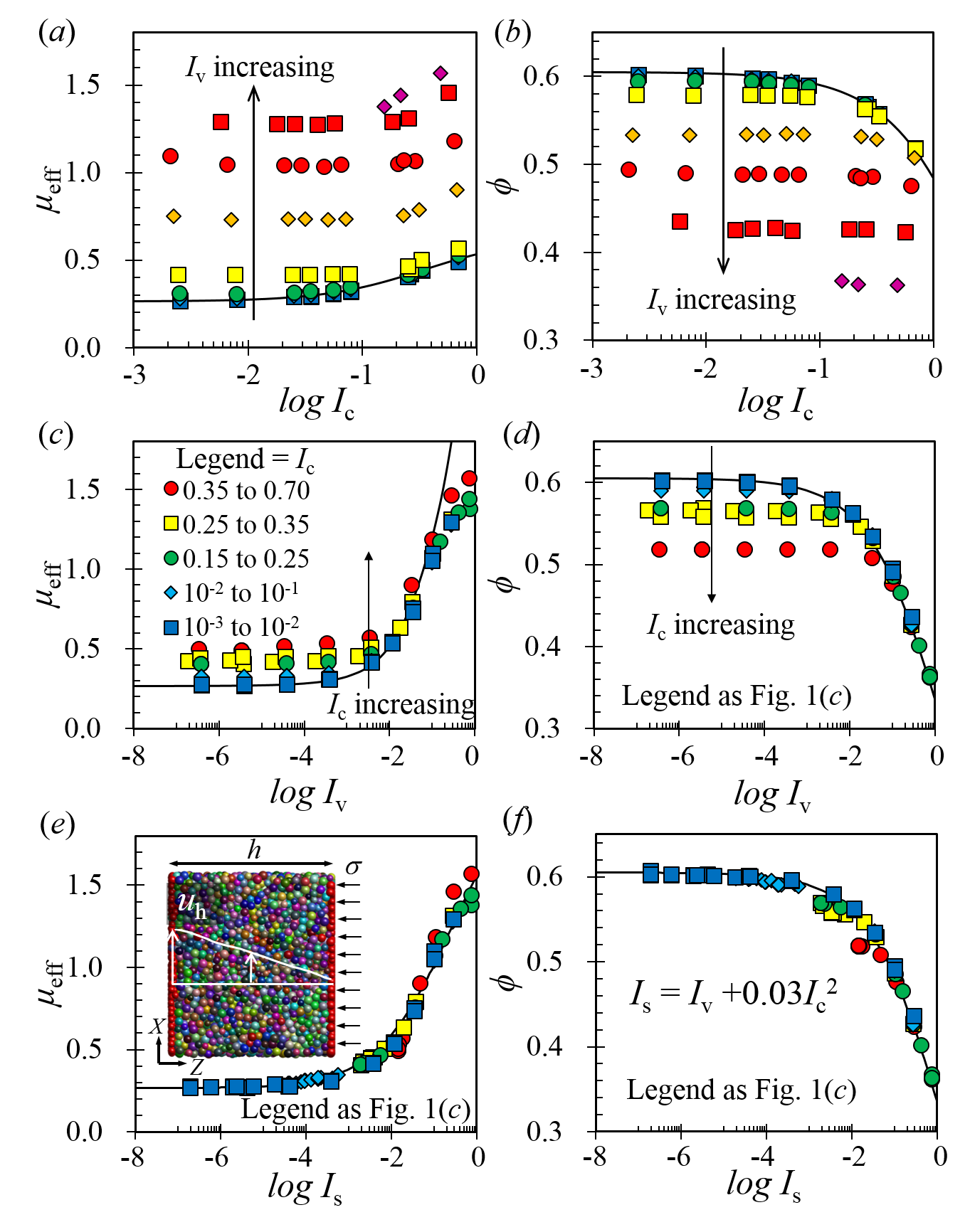}
  \caption{(a-b) plots of (a) $\mu_{\rm{eff}}$  vs.\ $I_c$ and (b) $\phi$ vs. $I_c$ for representative data sets. (a-b) Symbols indicate approximate $I_v$ values according to: \protect\tikzdiam{black}{purple} 0.74, \protect\tikzrect{black}{red} 0.28, \protect\tikzcirc{black}{red} 0.1, \protect\tikzdiam{black}{orange} 0.034, \protect\tikzrect{black}{yellow} $10^{-3}$ to $10^{-2}$, \protect\tikzcirc{black}{green} $10^{-4}$ to $10^{-3}$, \protect\tikzdiam{black}{lightblue} $10^{-5}$ to $10^{-4}$, \protect\tikzrect{black}{blue} $<10^{-5}$.  Solid lines show the best fit line of relationships for (a) $\mu_{\rm{eff}}$  and (b) $\phi$ suggested in Ref.\ \cite{jop2005} for dry spherical particles.  (c-d) plots of (c) $\mu_{\rm{eff}}$ vs.\ $I_v$ and (d) $\phi$ vs.\ $I_v$ for representative data sets.  (c-d)  Solid lines show the best fit line of relationships for (c) $\mu_{\rm{eff}}$  and (d) $\phi$ suggested in Ref.\ \cite{boyer2011}.  (e-f) plots of (e) $\mu_{\rm{eff}}$  and (f) $\phi$  for all simulated data. (e-f) Solid lines are the best fit relationships for $\mu_{\textrm{eff}}$  and $\phi$ suggested in Ref.\ \cite{trulsson2012} for slurries.   We obtain the best collapse when $\alpha \approx 0.03$, significantly smaller than that for 2-d slurries in Ref.\ \cite{trulsson2012} indicating  for our systems a higher sensitivity to viscous rather than collisional effects. (a-f) The equations and parameters for fits are provided in Table \ref{table:1}. (e) inset: sketch of simulations. The simulated cell size in the \textit{y-} and \textit{z-} directions are fixed at 30 mm and 58.3 mm, respectively; the steady state value for $h$ varies. }
  \label{Figure 1}
\end{figure}

To measure the rheology of these systems, we shear them in a rectangular box (Fig. \ref{Figure 1}) with periodic boundary conditions in the $x$- and $y$- directions and roughened walls (using ``glued'' particles) in the $z$-direction. We move one of the rough walls in the $x$-direction (only) with a constant velocity $u_h$. We apply a constant normal stress ($\sigma$) to the other roughened wall and allow it to move only in the $z$-direction. We run this simulation until the system reaches a statistically steady state and calculate the steady state average value of $h$.  We performed ten sets of simulations (see Supplemental Material II) designed to vary $I_c$ approximately from $10^{-3}$ to $1$ and $I_v$, when non-zero, from $10^{-7}$ to $1$. 

We find that our thickly-coated granular flows behave remarkably similarly to previously measured particle- and particle-fluid flows (Fig.\ \ref{Figure 1}): (1) for sufficiently low values of $I_v$ ($< 10^{-3}$), $\mu_{\rm{eff}}$ and $\phi$ are both essentially independent of $I_v$ and similar to those of dry granular flows \cite{jop2005}; (2) for sufficiently low values of $I_c$ ($< 10^{-1}$), $\mu_{\rm{eff}}$ and $\phi$ are both independent of $I_c$ and similar to those in suspensions and slurries \cite{boyer2011}; (3) for moderate-to-high values of $I_c$ and $I_v$,  $\mu_{\rm{eff}}$ and $\phi$ depend both on $I_c$ and $I_v$ and appear to collapse best with $I_s=I_v + \alpha I_c^2$. However, this is not as clean as a collapse as it first appears, apparent in a parametric plot of $\mu_{\rm{eff}}$ vs.\ $\phi$ (Fig.\ 2(a))

Rather than the monotonic relationship between $\mu_{\rm{eff}}$ and $\phi$, $\mu_{\rm{eff}}(\phi)$ is non-monotonic for these system in the range $0.51 \lessapprox \phi \lessapprox 0.61 \approx \phi_m$. $\phi \approx 0.51$ corresponds to the lower limit of $\phi$ for the highest value of $I_c$ at which our systems remain uniform. Within this range, $\mu_{\rm{eff}}$ is bounded at the low end with a predictive relationship for collisionally-dominated flows $\mu_{\rm{eff}}(\phi(I_c))$. The upper bound for $\mu_{\rm{eff}}$ appears to be dependent on both collisional and viscous effects $\mu_{\rm{eff}}(\phi(I_s))$ (Fig. 2(a) caption). We can transition systems from the lower limit curve $\mu_{\rm{eff}}(\phi(I_c))$ to the higher one ($\mu_{\rm{eff}}(\phi(I_s))$ with an increasing $I_v$ at constant $I_c$ (e.g., red arrow in Fig.\ (2(a) for $I_{c,max}$) by increasing $\eta$. With an initial increase in $\eta$, $\mu_{\rm{eff}}$ increases at a nearly constant $\phi$ to the $\mu_{\rm{eff}}(\phi(I_s))$ curve. If we increase $\eta$ further still, $\mu_{\rm{eff}}$ increases further while $\phi$ decreases along the $\mu_{\rm{eff}}(\phi(I_s))$ curve. 

\begin{figure}[h!]
  \includegraphics[scale = 0.35]{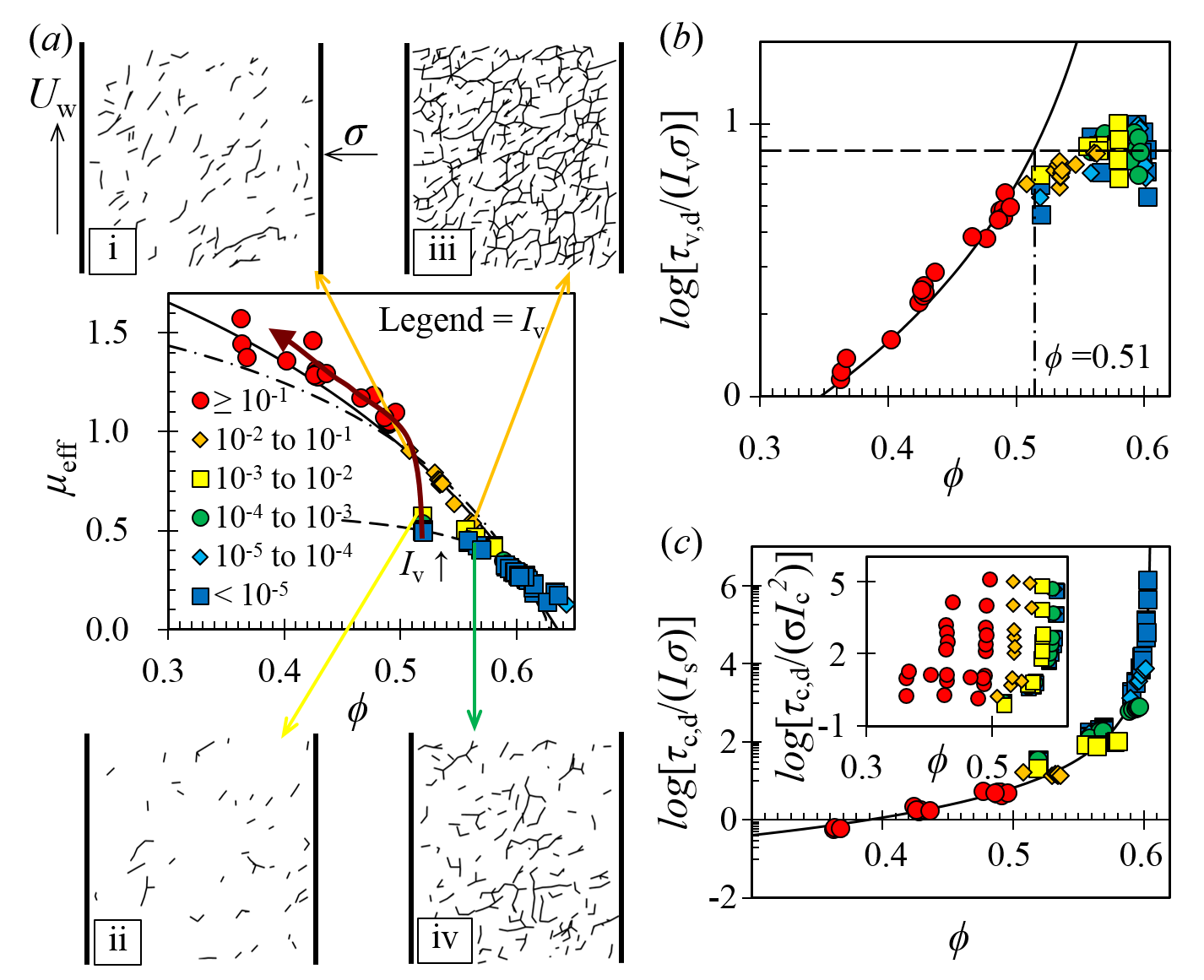}
  \caption{(a) main plot: parametric plot of $\mu_{\rm{eff}}$ vs.\ $\phi$ for all simulations performed herein.  The lines are: (dashed) $\mu_{\textrm{eff}} = \mu_{1c} + (\mu_{2c} - \mu_{1c}) / [1 + (\beta_{c} \times I_{co} \times \phi) / (\phi_m - \phi)]$ (line 1, Table 1); (dot-dahsed) $\mu_{\textrm{eff}} = \mu_{1s} + (\mu_{2s} - \mu_{1s}) / [1 + \beta_{s1}\sqrt{I_{so}}/(\phi_m - \phi)]$ line 3, Table 1; (solid) $\mu_{\textrm{eff}} = \mu_{1s} + (\mu_{2s} - \mu_{1s}) / [1 + \beta_{s1} \times \sqrt{I_{so}} \times \phi / (\phi_m - \phi)]$; (red arrow) constant-$I_c$ pathway for increasing $I_v$. (b) Ratio of computational to theoretical lubrication stresses $\tau_{\textrm{v,d}} / (\eta \dot{\gamma})= \tau_{\textrm{v,d}} / (I_v\sigma)$ vs. $\phi$.   solid line:  $\tau_{\textrm{v,d}}/(I_v\sigma) = (1/15)(\phi_m - \phi)^{-2}$; dashed line: $\tau_{\textrm{v,d}}/(I_v\sigma) = 8$, respectively.  (c) Ratio of computational contract stress to two different theoretical stresses.  inset:  $\tau_{\textrm{c,d}} /  [\alpha \rho_p (d \dot{\gamma})^2] = \tau_{\textrm{c,d}} / (I_c\sigma)$  vs.\ $\phi$; main plot: $\tau_{\textrm{c,d}} / [\alpha \rho_p (d \dot{\gamma})^2+ \eta \dot{\gamma}] = \tau_{\textrm{c,d}} / (I_s\sigma)$ vs.\ $\phi$. solid line: $\tau_{\textrm{c,d}} / (I_s\sigma) = (1/55)(\phi_m - \phi)^{-2.6}$. Insets to (a): instantaneous ``force diagrams'' from a slice of the shear cell ($13.5\ \textrm{mm} < y < 16.5\ \textrm{mm}$); each line connects centers of contacting particles for which the interparticle force $f_{ij}\geq \langle f_{ij} \rangle$, the mean force. [$I_c$, $I_v$] = (i) [0.68, 0.033] (ii) [0.69, 0.0034] (iii) [0.0075, 0.012] (iv) [0.25, 0.0037]. }
  \label{Figure 2}
\end{figure}

For a representation of force structures under these transitions, for two pairs of experiments with similar ($\phi$'s) and substantially different $\mu_{\rm{eff}}$'s we plot a representative ``strong force network'' -- line segments for particle pairs for which the interparticle contact force $f_{\rm{cij}}$ is greater than average (Fig.\ \ref{Figure 2}(a)).  For each pair on the same limit curve  $\mu_{\textrm{eff}}(\phi(I_c))$ or $\mu_{\textrm{eff}}(\phi(I_s))$, increasing $\mu_{\rm{eff}}$ simultaneously decreases $\phi$ and both connectivity and density of high$-f_{\rm{cij}}$ force pairs. At the same time, enduring extended frictional contact networks are replaced by more isolated collisional contacts. In contrast, for each pair with similar $\phi$'s, increasing $\mu_{\rm{eff}}$ limit curve \textit{increases} both connectivity and density of high$-f_{\rm{cij}}$ force pairs. Based on these observations, we hypothesize that as a system transitions from the low limit $\mu_{\rm{eff}}$ curve to the high limit curve increasing $I_v$ at constant $I_c$, modest increases in viscous damping of interparticle movement decreases separation events of contacting particles, similar to an effective ``stickiness'', increasing relative strong contact force connectivity. This results in an increase in $\tau$ relative to $\sigma$ without much change in $\phi$. Once the upper $\mu_{\rm{eff}}$ limit curve is reached, further increases in $I_v$ reduces particle contact events transitioning the system back to one where interparticle contacts are more isolated and collisional.

From our data, we calculate a contact and viscous shear stresses ($\tau_{c,d}$ and $\tau_{v,d}$) from the $x$-component of all contact forces and viscous forces between each wall particle and adjacent free particles. We find that $\tau_{v,d} / \tau_v = \tau_{v,d} / (\eta \dot{\gamma})$ can be expressed as a single-valued function of $\phi$, $f_v(\phi)$ (Fig.\ 2(b)) though $\tau_{c,d}  / \tau_c = \tau_{c,d} / [\rho_p (d \dot \gamma)^2]$ cannot (Fig.\ 2(c), inset). Rather, $\tau_{c,d} / [\alpha \rho_p (d \dot \gamma)^2 + \eta\dot{\gamma})]$ is much closer to a single-valued function of $\phi$, $f_{\textrm{c}}(\phi$) (Fig.\ 2(c) and caption).  We note that this $\phi$-dependent behavior suggest that, in these systems, while non-contact viscous interactions are influenced by $I_v$ alone, contact interactions are influenced by both $I_c$ and $I_v$. Considering the significantly different forms of $f_v(\phi)$ and $f_c(\phi)$, we suggest that in place of Eqn.\ (2-3) for thickly-coated systems we write:
\begin{subequations} \label{eq:5}
\begin{align}
\tau=f_{\textrm{c}}(\phi)\times [\alpha \rho_p (d \dot\gamma)^2 +\eta\dot\gamma] + f_{\textrm{v}}(\phi)\times \eta\dot\gamma \label{eq:5a} \\
\mu_{\rm{eff}}=f_{\textrm{c}}(\phi)\times \alpha I_c^2 + [f_{\textrm{c}}(\phi)+ f_{\textrm{v}}(\phi)]\times I_v \label{eq:5b}
\end{align}
\end{subequations}
Equation \ref{eq:5}(b) provides an effective, if empirical, expression for the data in Fig.\ 2(a).  Further, in certain limits and with Figs.\ 2(b-c) it provides intuition to the structure of $\mu_{\rm{eff}}(\phi)$. For $\phi \gtrapprox 0.56$, $f_{\rm{c}} / f_{\rm{v}} \gg 1$, so for these data $\mu_{\rm{eff}} \approx  f_{\textrm{c}}(\phi)\times I_s$ signifying a regime where viscous and contact effects influence the dynamics.  For $\phi \lessapprox 0.56$ all data on the $\mu_{\rm{eff}}(\phi(I_s))$ (upper bound) curve in Fig.\ 2(a) correspond to cases where $1 \ll  \alpha I_c^2 / I_v \equiv St$ (a Stokes number). For these data $\mu_{\rm{eff}}\approx [f_{\textrm{c}}(\phi)+ f_{\textrm{v}}(\phi)]\times I_v$ signifying a regime where viscous effects dominate systems on the upper $\mu_{\rm{eff}}$ curve for $\phi \lessapprox 0.56$.  While the functional forms for this upper bound $\mu_{\rm{eff}}$ curve differ, at $\phi \approx 0.56$, $I_v \approx I_s$ and $f_{\textrm{c}}(\phi)+ f_{\textrm{v}}(\phi) \approx f_{\textrm{c}}(\phi)$, so there is a smooth transition. Now considering the lower  bound $\mu_{\rm{eff}}=\mu_{\rm{eff}}(\phi(I_c))$ curve: for $\phi \lessapprox 0.56$, $St \gg 1$ and $f_{\textrm{c}}(\phi) \approx f_{\textrm{v}}(\phi)$ so $\mu_{\rm{eff}} \approx f_{\rm{c}}(\phi) \times \alpha I_c^2$, dependent primarily on contact effects, as Fig.\ \ref{Figure 2}(a) and caption imply. For larger $\phi$'s on this curve, $I_c, St$ decrease and the functional form $\mu_{\rm{eff}}=f_{\textrm{c}}(\phi)\times I_s$, the same form as the upper bound curve in this region of $\phi$.   

\begin{figure}[h!]
  \includegraphics[scale = 0.35]{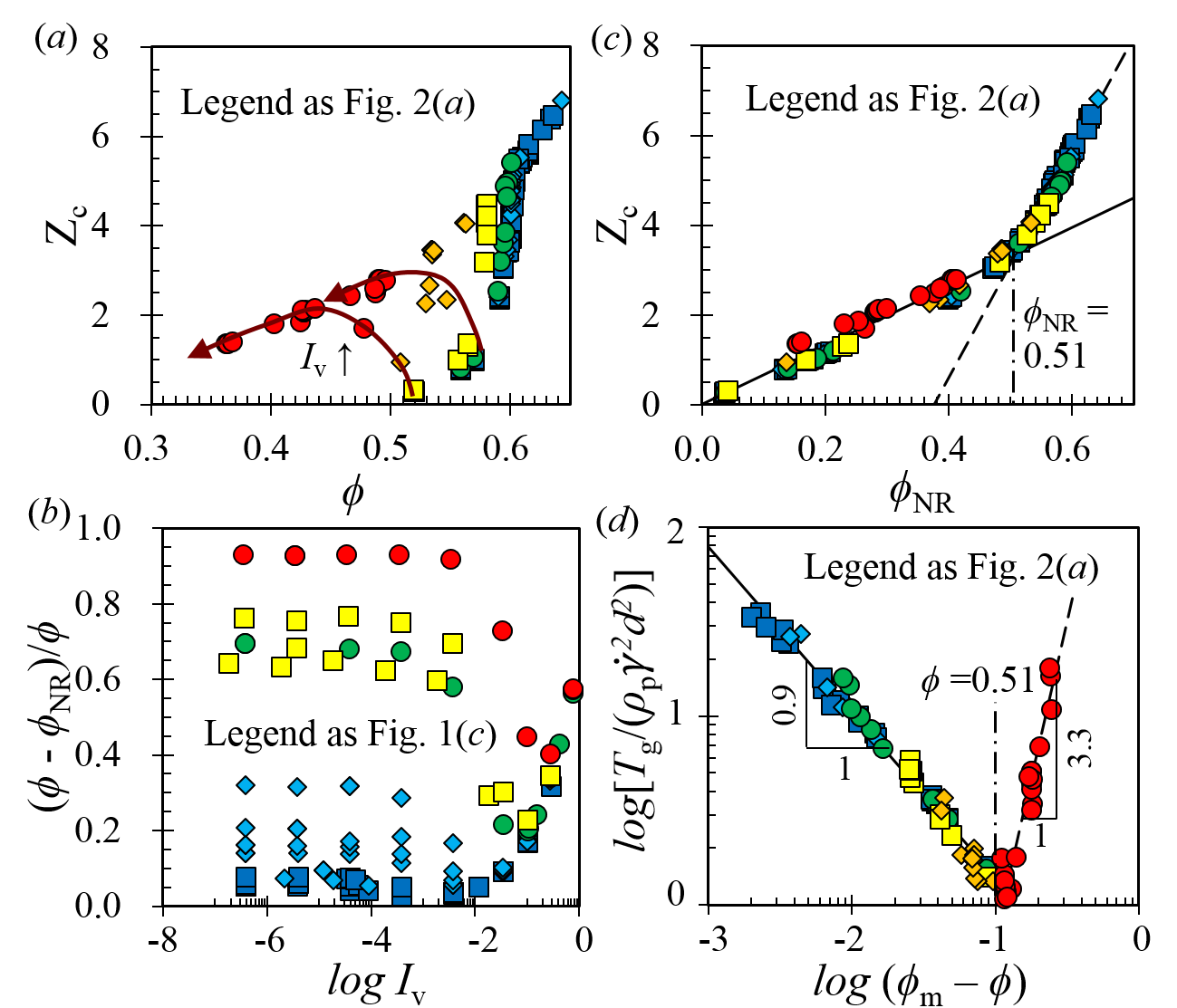}
  \caption{(a) Coordination number $Z_c$ plotted vs.\ $\phi$. red arrows: pathways of increasing $I_v$ for constant $I_c$. (b) $F$, the fraction of particles with $<2$ contacts plotted vs.\ $I_v$; symbols indicate value of $I_c$. (c) $Z_c$ vs.\ $\phi_{\rm{NR}}$: $Z_c = 6.6\phi_{\rm{NR}}$; Dashed curve: $Z_c = 25.3\phi_{\rm{NR}} - 9.5$. (d) Normalized granular temperature, $T_{g}/(\rho_{p}\dot{\gamma}^2 d^2)$ plotted vs.\ log($\phi-\phi_m$), ($\phi_m$=0.61).}
  \label{Figure 3}
\end{figure}

This general picture is supported by consideration of the relationship between \textit{coordination number} $Z_c$ (average number of contacts per particle), $\phi$, and other bulk parameters (Figs.\ \ref{Figure 3}(a-c)). As indicated in Fig.\ \ref{Figure 3}(a) (dark red arrows), for the same moderate increase in $I_v$ for which the system remained primarily on the low $\mu_{\rm{eff}}(\phi(I_c))$ limit curve, $Z_c$ maintains a near-constant small value. We also check the fraction of particles potentially available for a connected strong force network (Fig.\ \ref{Figure 3}(b)); $\phi_{\rm{NR}}$ is the part of the system solid fraction comprised of particles contacting more than one particle (NR stands for non-rattler particles, as in Ref.\ \cite{bi2011jamming})).  For highest $I_c$ and $I_v<0.2$, $F \equiv (\phi-\phi_{\rm{NR}})/\phi \approx 0.93$ indicating that on average only $\approx 7\%$ of the particles have more than one contact, supporting the picture of a system dominated by occasional collisions.  Then $F$ first decreases as $Z_c$ increases (from $I_v \approx$ 0.2 to 0.1), signifying the more highly connected force network.  With further $I_v$ increase, $F$ rises as $Z_c$ drops again, signifying the return to a less well-connected contact network. We find when we plot $Z_c$ vs.\ $\phi_{\rm{NR}}$, we get a convincing collapse, suggesting as others have found for somewhat different particulate systems \cite{bi2011jamming}, $Z_c$ and $\phi_NR$ are much more deterministic of system behavior than $\phi$.   

To summarize, we find that thickly-coated particle-fluid flows behave in many ways similarly to fully saturated particle-fluid flows, particularly in their response to changing either viscous or contact effects constant. However, simultaneous variation of both effects reveals concentration-dependent transitions in bulk behaviors reflected in interparticle forces and force networks not previously reported for similar particle-fluid systems.  Ongoing work includes investigations of the transitions between particle-scale interactions of these systems and dynamics in analogous fully-saturated systems.

We finish with two additional observations. First, somewhat unexpectedly to us, the \textit{granular temperature} $T_g \sim \rho_s \phi \langle (\vec{u} - \langle \vec{u} \rangle)^2 \rangle$ plotted vs.\ $\phi_m-\phi$ does not help us distinguish between collisional and visco-collisional flows. Rather, we for nearly all of our systems, $T_g$ has the same functional dependence on $\phi_m-\phi$, not dissimilar from that predicted by Bagnold for moderate-to-dense granular-fluid flows \cite{bagnold1954} (see supplement). The exceptions we found (not surprisingly) were for $\phi \lessapprox 0.51$ and $I_v>0.1$ for which we found $T_g$ increases with increasing $\phi_m-\phi$ (Fig.\ 3(c)). Finally, for future refreence we note the wealth of related studies at the ``jamming'' (liquid-to-solid) transition from which we can seek additional insight for transitions in our flowing systems. In their 2-d experimental jamming studies, Bi et al.\ \cite{bi2011jamming} found a similar functional form and slope change in $Z_c$ vs.\ $\phi_{\rm{NR}}$ as we did in our flowing system (Fig.\ 3(c)) as well as a slope change with a change of fabric structures. Links like this gives insight to a more broadly unified physical framework for dense particle-fluid deformation and flows.   

We gratefully acknowledge the funding for this research provided by the NSF under the grant EAR-1451957 on "Entrainment and Deposition of Surface Material by Particle-Laden Flows", the UMN Center of Transportation Studies and the CEGE Sommereld Fellowship and computing resources provided by SAFL at UMN. The authors also thank Prof.\ Jia-Liang Le for helpful discussions. 
\bibliographystyle{apsrev4-1}
\bibliography{Rheo_ref}
\end{document}